\documentclass[aps,prl,reprint,twocolumn,noshowpacs,superscriptaddress]{revtex4-2}  

\usepackage[utf8]{inputenc}
\usepackage{graphicx}
\usepackage{amsmath}
\usepackage{amssymb}
\usepackage{comment}
\usepackage{bm}        
\usepackage{txfonts}
\usepackage{multirow}
\usepackage{siunitx}
\usepackage{xargs}                      
\usepackage{microtype}

\usepackage[pdftex,dvipsnames]{xcolor}  
\usepackage[normalem]{ulem} 

\usepackage[%
  colorlinks=true,
  urlcolor=blue,
  linkcolor=blue,
  citecolor=blue
]{hyperref}

\usepackage{SM_theory_notation}

\begin{document}

\title{High-purity quantum optomechanics at room temperature}

\author{Lorenzo Dania}
\affiliation{Photonics Laboratory, ETH Z\"urich, 8093 Z\"urich, Switzerland}
\affiliation{Quantum Center, ETH Z\"urich, 8093 Z\"urich, Switzerland}
\author{Oscar Schmitt Kremer}
\affiliation{Photonics Laboratory, ETH Z\"urich, 8093 Z\"urich, Switzerland}
\affiliation{Quantum Center, ETH Z\"urich, 8093 Z\"urich, Switzerland}
\author{Johannes Piotrowski}
\affiliation{Photonics Laboratory, ETH Z\"urich, 8093 Z\"urich, Switzerland}
\affiliation{Quantum Center, ETH Z\"urich, 8093 Z\"urich, Switzerland}
\author{Davide Candoli}
\affiliation{ICFO – Institut de Ciencies Fotoniques, The Barcelona Institute
of Science and Technology, Castelldefels, Barcelona 08860, Spain}
\author{Jayadev Vijayan}
\affiliation{Photonics Laboratory, ETH Z\"urich, 8093 Z\"urich, Switzerland}
\affiliation{Quantum Center, ETH Z\"urich, 8093 Z\"urich, Switzerland}
\affiliation{Photon Science Institute, Department of Electrical and Electronic Engineering, University of Manchester, Manchester, UK}
\author{Oriol Romero-Isart}
\affiliation{ICFO – Institut de Ciencies Fotoniques, The Barcelona Institute
of Science and Technology, Castelldefels, Barcelona 08860, Spain}
\affiliation{ICREA – Institucio Catalana de Recerca i Estudis Avancats, Barcelona 08010, Spain}
\author{Carlos Gonzalez-Ballestero}
\affiliation{Institute for Theoretical Physics, Vienna University of Technology (TU Wien), 1040 Vienna, Austria}
\author{Lukas Novotny}
\affiliation{Photonics Laboratory, ETH Z\"urich, 8093 Z\"urich, Switzerland}
\affiliation{Quantum Center, ETH Z\"urich, 8093 Z\"urich, Switzerland}
\author{Martin Frimmer}
\affiliation{Photonics Laboratory, ETH Z\"urich, 8093 Z\"urich, Switzerland}
\affiliation{Quantum Center, ETH Z\"urich, 8093 Z\"urich, Switzerland}

\date{\today}

\begin{abstract}
Exploiting quantum effects of mechanical motion, such as backaction evading measurements or squeezing, requires preparation of the oscillator in a high-purity state. The largest state purities in optomechanics to date have relied on cryogenic cooling, combined with coupling to electromagnetic resonators driven with a coherent radiation field. 
In this work, we cool the mega-hertz-frequency librational mode of an optically levitated silica nanoparticle from room temperature to its quantum ground state.
Cooling is realized by coherent scattering into a Fabry-Perot cavity.
We use sideband thermometry to infer a phonon population of 0.04 quanta under optimal conditions, corresponding to a state purity of $92\%$. 
The purity reached by our room-temperature experiment exceeds the performance offered by mechanically clamped oscillators in a cryogenic environment.
Our work establishes a platform for high-purity quantum optomechanics at room temperature.

\end{abstract}
\maketitle

\textit{Introduction}. 
The prospect of observing and exploiting quantum states of massive systems has been driving the field of optomechanics~\cite{Aspelmeyer2014}. 
Mechanical motion controlled by optical or microwave fields offers opportunities to develop quantum-enhanced sensing schemes~\cite{LecocqQND_2015,Shomroni-BAE_2019,Barzanjeh_2021} and transduction technologies~\cite{AndrewsConversion_2014},
test quantum mechanics at unprecedented mass and length scales~\cite{arndt2014testing,Croquette_Mesoscopic2023}, and gain insights into the role of the gravitational field in the evolution of quantum states~\cite{Rickles1957chapelHill,Belenchia2018}.
Crucial to these applications is the suppression of thermal noise, which calls for operation in a cryogenic environment.
While highly effective, cost and technical complexity of cryogenic cooling severely limit further proliferation of optomechanical technologies. 
The promise of room-temperature quantum optomechanics has therefore spurred the development of experimental platforms operating without the need for cryogenic cooling~\cite{Guo-FeedbackCooling_2019}.

The necessity to suppress thermal noise is rooted in the requirement of any quantum protocol to initialize the oscillator in a quantum mechanically pure state~\cite{TendickQuantifying_2022,roda-llordes2023macroscopic}. 
For a thermal state with mean occupation number $n$, the purity is given by $\mathcal{P}=(2 n +1)^{-1}$~\cite{Paris-Purity_2003}.
For sufficiently large mechanical frequencies, like giga-hertz mechanical modes of nanobeams~\cite{Riedinger2016} or bulk acoustic-wave resonators~\cite{chu2017}, state purification can be achieved by thermalization to a cryogenic bath sufficiently cold to render the oscillator in its ground state of motion.
For mega-hertz or even lower mechanical frequencies,  high-purity state preparation requires a combination of cryogenic cooling with additional techniques like cavity sideband cooling~\cite{Youssefi2023,chan2011laser,peterson_cavity_radiation_pressure} or measurement-based feedback cooling~\cite{RossiMeasurement_2018,tebbenjohanns2021quantum}.
With such combined cooling schemes, mega-hertz electro-mechanical oscillators have been prepared with a phonon occupation of 0.07 (88\% state purity)~\cite{Youssefi2023} and giga-hertz optomechanical systems have been brought to a purity of 85\%~\cite{Qiu_laserCooling-2020}.

The key to circumvent the need for cryogenic cooling is to suppress the coupling of the mechanics to its thermal environment. 
In this vein, two approaches have been followed. The first is based on complete mechanical decoupling of the oscillator from its environment by optical levitation in vacuum~\cite{gonzalezballestero2021levitodynamics}. 
Cooling the center-of-mass motion of an optically levitated nanoparticle in room-temperature experiments has been reported using both measurement-based feedback~\cite{magrini2020} and laser-sideband cooling ~\cite{Delic2020}, reaching a phonon population of 0.6, corresponding to a state purity of $47\%$~\cite{magrini2020}.
The second approach focuses on careful design of the strain and phononic dispersion of the mechanical tether~\cite{maccabe2020,Beccari2022}.   
In a tour-de-force of mechanical and optical engineering, recent efforts have enabled quantum optomechanics at room temperature with a clamped system, reaching a purity of 34\%~\cite{Huang_2024}. Despite these efforts, no room-temperature optomechanical platform can currently rival the state purities achieved with the aid of cryogenics.

Here, we report cooling of a mega-hertz librational mode of an optically levitated nanoparticle to a phonon population of $n =0.04$ in a room-temperature experiment. Cooling is realized by coupling the nanoparticle to a high-finesse optical cavity in a coherent-scattering configuration~\cite{stickler_elliptic_coherent}. The achieved state purity reaches $\mathcal{P}=92\%$, enabled by active suppression of laser phase noise.
Regarding quantum mechanical purity, our results place levitated oscillators in room-temperature experiments ahead of the most performant opto- and electromechanical systems, even those aided by cryogenics and giga-hertz mechanical mode frequencies.

\begin{figure}[htb]
\includegraphics[width=1\columnwidth]{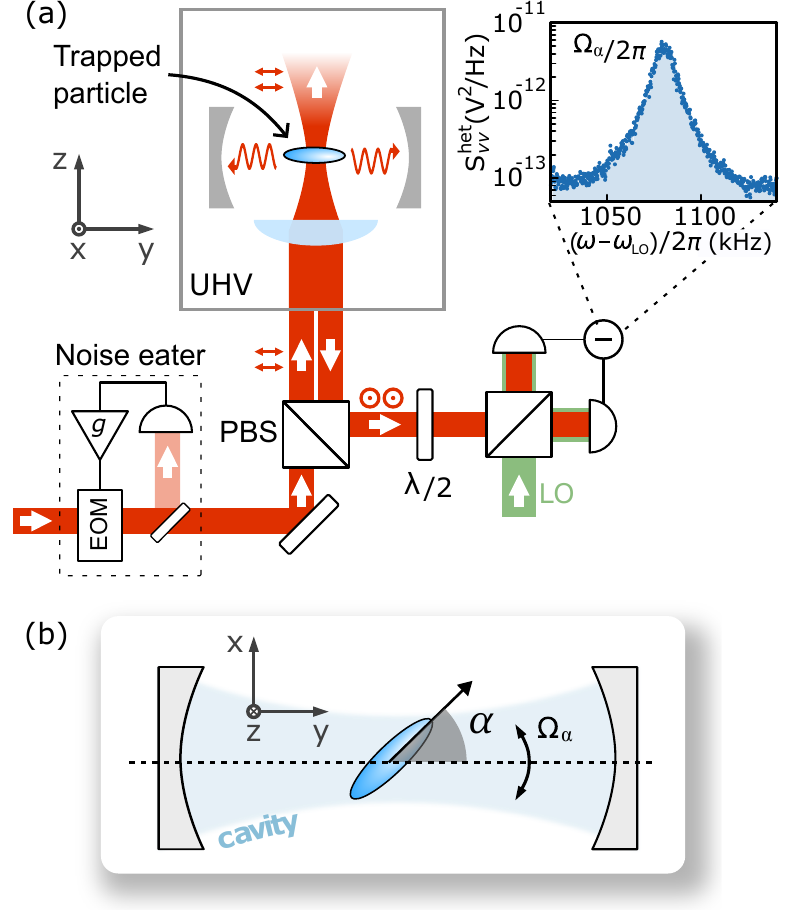}
\caption{\textbf{Sketch of the experimental setup.}
(a)~An anisotropic silica nanoparticle (schematically illustrated as an ellipsoid) is trapped by an optical tweezer in ultra-high vacuum (UHV). 
The tweezer light is linearly polarized along the $y$ axis by a polarizing beam splitter (PBS).
The long axis of the nanoparticle aligns parallel to the tweezer polarization and undergoes angular harmonic oscillations, termed \textit{librations}, at a frequency $\Omega_\alpha/(2\pi)$ in the $xy$ plane.
This libration motion is coupled to a high-finesse optical cavity.
The high-NA lens forming the optical trap is mounted on a nanopositioner (not shown), such that the particle equilibrium position can  be varied across the cavity intensity profile.
The $x$-polarized back-scattered light from the  nanoparticle is collected by the trapping lens and mixed with a local oscillator of frequency $\omega_{\textrm{LO}}$ in a balanced heterodyne detector. 
This detector provides a measurement of the libration motion unaltered by the cavity transfer function.
The inset shows a power spectral density $S_{VV}^{\textrm{het}}$ from this backward detector with the particle’s librational mode peaked at $\Omega_\alpha/(2\pi)=\SI{1.08}{\mega\hertz}$.
Laser phase noise in the tweezer beam can be suppressed by a noise eater composed of a phase noise detector and an electro-optic modulator (EOM). The suppression level is varied with a gain $g$.  
(b)~Illustration of libration mode. 
The tweezer polarization is aligned to the cavity axis ($y$).
The libration angle $\alpha$ denotes the deviation of the particle's long axis from the polarization direction of the tweezer field in the $xy$ plane. 
}
\label{fig:fig1}
\end{figure}
\textit{Experimental platform.} 
The nanomechanical oscillator investigated in this work is the harmonic angular motion of an optically-levitated anisotropic particle~\cite{hoang2016torsional}. 
An anisotropic scatterer in a linearly polarized field aligns its axis of largest polarizability with the field's polarization direction~\cite{bang2020five}. Small deviations from this alignment result in harmonic angular motion, termed \textit{libration}~\cite{Stickler2021quantumrotations}.
Figure~\ref{fig:fig1}(a) shows a sketch of the experimental setup, with more details given in the Supplemental Material~\cite{supp_mat_libration}.
We use an optical tweezer (power $\sim \SI{1.2}{\watt}$, numerical aperture NA = 0.75, wavelength $\lambda=\SI{1550}{\nano\meter}$) to trap single anisotropic nanoparticles inside a vacuum chamber at a pressure of \SI{5e-9}{\milli\bar} and at room temperature.
The tweezer beam propagates along the $z$ direction and is linearly polarized along the $y$ axis, resulting in center-of-mass frequencies $(\Omega_{x},\Omega_{y},\Omega_{z})/(2\pi)=(250,220,80)\,$\si{\kilo\hertz}.
We trap anisotropic nanoparticles, which are clusters of few silica nanospheres with a nominal diameter of \SI{120}{\nano\meter}, each.
The measured center-of-mass gas damping rates~\cite{supp_mat_libration} indicate that the nanoparticle has its long axis aligned to the tweezer polarization~\cite{Hoang2016}, and that its shape is not cylindrically symmetric~\cite{gao2024}.
We detect the orientation of the nanoparticle by interfering backscattered light from the tweezer with a local oscillator shifted by $\omega_{\mathrm{LO}}/(2\pi)=\SI{2.73}{\mega\hertz}$ in a balanced heterodyne scheme. 
The optical tweezer is positioned in the waist of a high-finesse optical cavity whose axis is oriented perpendicularly to the tweezer axis.
Light scattered by the nanoparticle populates the fundamental TEM00 mode of the cavity, resulting in optomechanical coupling via coherent scattering~\cite{supp_mat_libration}.
The cavity mode has a linewidth $\kappa/(2\pi) = \SI{330}{\kilo\hertz}$ and a resonance frequency $\omega_{\mathrm{c}}=\omega_{\mathrm{tw}}+~\Delta$, detuned by $\Delta$ from the tweezer frequency $\omega_{\mathrm{tw}}$.
The nanoparticle position $y_{\mathrm{eq}}$ along the cavity standing wave is tunable via a nanopositioner holding the trapping lens. 
Finally, our system contains a noise-eater that allows us to controllably suppress laser-phase noise~\cite{ivan2021}.

The anisotropic particle shape results in three distinct moments of inertia~\cite{supp_mat_libration}, giving rise to three non-degenerate libration modes associated with the orientation angles $\gamma$, $\beta$, and $\alpha$~\cite{Pontin2023,kamba2023}, with corresponding frequencies $(\Omega_{\gamma},\Omega_{\beta},\Omega_{\alpha})/(2\pi)=(0.15,0.7,1.08)$~\si{\mega\hertz}.
The high-frequency $\alpha$ mode corresponds to angular oscillations in the tweezer focal plane [$xy$ plane in Fig.~\ref{fig:fig1}(b)]~\cite{gao2024}. The inset of Fig.~\ref{fig:fig1}(a) shows a high-pressure ($\SI{6}{\milli\bar}$) spectrum of the heterodyne signal. We associate the Lorentzian peak with the motion of the $\alpha$ mode.
In this work, we focus on cavity-cooling this mode  for two reasons:
First, the $\alpha$ mode lies deep in the sideband-resolved regime ($\Omega_{\alpha}\gg\kappa$), which is a required condition to reach the ground state via cavity cooling~\cite{Schliesser2008}. 
Second, optimal cooling of the $\alpha$ mode is achieved when polarizing the tweezer along the $y$ axis [see Fig.~\ref{fig:fig1}(b)]~\cite{stickler_elliptic_coherent}. At the same time, this configuration directs the dipole radiation pattern of light elastically scattered by the nanoparticle outside of the cavity, thus minimizing heating effects of laser phase noise~\cite{meyer2019}.
As the particle orientation oscillates in the $xy$ plane, light from the tweezer gets inelastically scattered into the cavity mode at the Stokes and anti-Stokes frequencies $\omega_{\mathrm{tw}}\pm \Omega_{\alpha}$.
For a cavity detuning of $\Delta\approx\Omega_{\alpha}$, the anti-Stokes scattering is enhanced, promoting energy transfer from the mechanics to the light field and resulting in cavity cooling of the librational motion.

In the following, we detail how we maximize state purity of our levitated librator. To this end, we first benchmark our thermometry scheme. Then, we optimize the cavity detuning for best cooling performance. Finally, we optimize the particle position in the cavity mode in the presence of phase-noise suppression.

\begin{figure}[ht]
\includegraphics[width=\columnwidth]{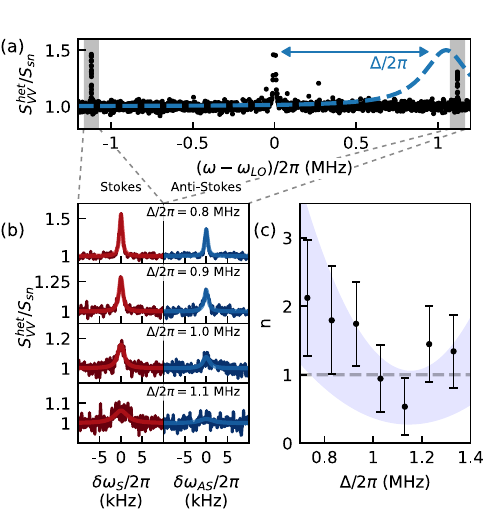}
\caption{
\textbf{Libration occupation number as a function of cavity detuning.} 
(a)~Heterodyne libration spectrum measured with the free-space backward detector (black), superposed with the cavity lineshape (blue dashed line). The detuning $\Delta$ is the difference between the tweezer and the cavity resonance frequency.
Highlighted in gray are the Stokes and anti-Stokes librational sidebands.
For $\Delta\approx\Omega_\alpha$, with $\Omega_\alpha/(2\pi)=\SI{1.08}{\mega\hertz}$ being the librational frequency, the cavity enhances the anti-Stokes scattering, leading to resolved sideband cooling.
(b)~Stokes (left) and anti-Stokes (right) heterodyne spectra $S_{VV}^\text{het}$ normalized to the shot-noise level $S_\text{sn}$ with fitted Lorentzian lines for $\Delta$ increasing from top to bottom as indicated in the plot. 
$\delta\omega_\mathrm{S}$ and $\delta\omega_\mathrm{aS}$ denote the frequency difference from the Stokes and anti-Stokes peak center, respectively.
(c)~Occupation number $n$ obtained via sideband thermometry as a function of $\Delta$ reaching  a minimum of $n=0.5(3)$ at $\Delta /(2\pi) = \SI{1.13}{\mega\Hz}$. 
Error bars are standard deviations of fitted parameters. The shaded area corresponds to a theoretical estimation of $n$ based on libration-cavity coupling, laser phase noise, radiation torque shot noise and their uncertainties (see Supplement~\cite{supp_mat_libration}).
}
\label{fig:fig2}
\end{figure}
\textit{Sideband thermometry and cavity detuning scan.} 
The free-space heterodyne detector gives us access to both Stokes and anti-Stokes sidebands of the $\alpha$ librational mode.
We verify cavity cooling via Raman sideband thermometry~\cite{magrini2020, tebbenjohanns2021quantum}.
Figure~\ref{fig:fig2}(a) shows a heterodyne spectrum of the libration motion (black dots, normalized to the shot noise level) obtained for cavity detuning $\Delta/(2\pi) = \SI{0.8}{\mega\hertz}$.
The plot shows the Stokes (left) and anti-Stokes (right) libration peaks, highlighted by the shaded gray bands.
The area below the Stokes sideband $a_{\mathrm{S}}$ (anti-Stokes sideband $a_{\mathrm{aS}}$) is associated with light-scattering events that increase (decrease) the librational energy. 
An asymmetry between the two sidebands arises as the particle's angular motion gets cooled close to its ground state and we can
deduce the libration occupation number $n$ from the ratio of the sidebands $a_{\mathrm{aS}}/a_{\mathrm{S}}=n/(n+1)$~\cite{clerk2010}.
We stress that, contrary to schemes detecting the cavity output spectrum~\cite{peterson_cavity_radiation_pressure,piotrowski2023simultaneous,Delic2020}, our free-space detection 
does not rely on a prior knowledge of the cavity detuning, and it is insensitive to sideband artifacts from classical laser phase noise which may corrupt the occupation estimation~\cite{Jayich_2012,Safavi-Naeini_2013}. Furthermore, we rule out the detector transfer function as a possible source of sideband asymmetry in the Supplemental Material~\cite{supp_mat_libration}.

By using sideband thermometry, we measure the libration occupation number as a function of the cavity detuning. For this measurement, the particle is positioned at $ky_{\mathrm{eq}} \approx 0.1\pi$ in the cavity standing wave, where $k$ is the wave number, and we define the position  along the cavity axis such that $y_{\mathrm{eq}}=0$ coincides with the intra-cavity intensity minimum.  
Figure~\ref{fig:fig2}(b) shows heterodyne spectra normalized to the detection shot-noise level $S_{\mathrm{sn}}$
centered around the Stokes (left column) and anti-Stokes (right column) libration peaks, and for increasing values of the cavity detuning $\Delta$ (top to bottom).
For better comparison, the spectra have been shifted along the frequency axis to align the Stokes and anti-Stokes peaks, respectively. The original spectra are shown in the Supplemental Material~\cite{supp_mat_libration}.
As the detuning approaches the optimal value $\Delta\approx\Omega_{\alpha}\approx2\pi\times\SI{1.1}{\mega\hertz}$, the linewidth increases and the peak height decreases due to cavity cooling.
From a Lorentzian fit [lines in Fig.~\ref{fig:fig2}(b)] to each lineshape we extract the occupation number $n$.
In Fig.~\ref{fig:fig2}(c) we plot the measured occupation number $n$ including the standard deviation of the fits as function of the cavity detuning $\Delta$.
For $\Delta  = 2\pi\times\SI{1.1}{\mega\Hz}$ we obtain the minimum occupation of $n=0.5(3)$.
The blue area in Fig.~\ref{fig:fig2}(c) is a theory estimation based on our system parameters, including the libration-cavity coupling rate, laser phase noise of the tweezer beam, and radiation torque shot noise~\cite{supp_mat_libration}.
Our model suggests that the final occupation number in this experiment is limited by laser phase noise in the tweezer beam, a limitation that has plagued already previous optomechanics experiments~\cite{nadine_phase}. 
\begin{figure*}[ht]
\includegraphics[width=2\columnwidth]{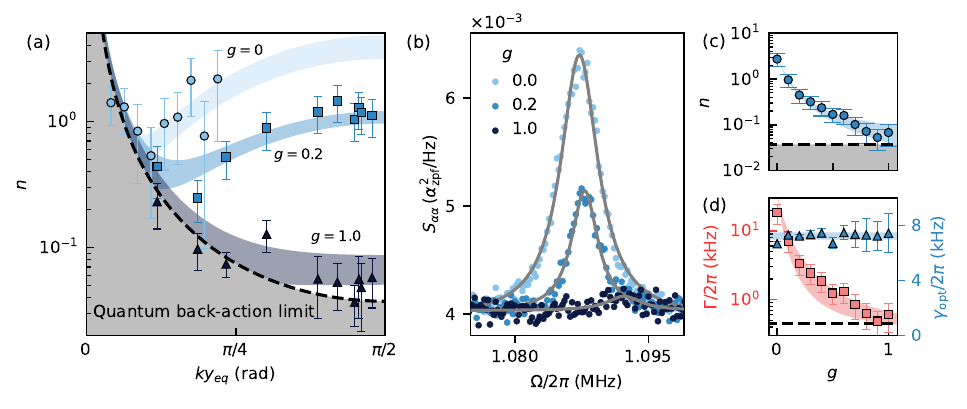}
\caption{\textbf{Cavity cooling dependence on particle position in the standing wave and phase noise}.
(a)~Occupation number $n$ measured for different particle positions $y_{\mathrm{eq}}$ along the cavity standing wave and for different levels of phase-noise cancellation $g$ as indicated.
The cavity intensity minimum coincides with the position $y_{\mathrm{eq}} = 0$ (cavity node), while the maximum occurs for $ky_{\mathrm{eq}} = \pi/2$ (cavity anti-node).
The colored areas are fits to our model including phase noise~\cite{supp_mat_libration}.
The black dashed line corresponds to the quantum back-action limit reachable in the absence of phase noise. (b) Homodyne spectra $S_{\alpha \alpha}$ of the libration degree of freedom for different cancellation gains and with the particle positioned at the anti-node. Gray solid lines are Lorentzian fits. (c)~Occupation number $n$ as a function of phase-noise cancellation $g$ for a particle placed in the cavity antinode at optimal detuning. The blue area represents the theory prediction, while the black dashed line is the quantum back-action limit. 
(d)~Total heating rate $\Gamma$ (red) and optical cooling rate $\gamma_{\mathrm{opt}}$ (blue) as a function of $g$, extracted from the same dataset underlying (b). The colored areas following the data points 
are theory predictions based on our model. The black dashed line is the heating rate due to radiation torque shot noise extracted from the fits in (b). 
}
\label{fig:fig3}
\end{figure*}\\
\textit{Position dependence of cooling performance.} 
To corroborate our understanding, we study the dependence of the cooling performance as a function of particle position in the cavity. While heating due to laser phase noise is position dependent~\cite{meyer2019,Delic2020}, radiation torque shot noise and heating from gas collisions~\cite{van2021sub} do not depend on position.
The blue circles in Fig.~\ref{fig:fig3}(a) show the occupation $n$ measured for different positions $y_{\mathrm{eq}}$ along the standing wave of the cavity field, whose periodicity is set by the wavenumber $k$.
This experiment is done at optimal cavity detuning $\Delta\approx\Omega_{\alpha}$.
%
We observe that the occupation reaches a minimum value of $n=0.5$ for $ky_{\mathrm{eq}} \approx 0.1\pi$. This is the position where the data in Fig.~\ref{fig:fig2}(c) has been acquired.
The initially decreasing trend of $n$ with position $y_\mathrm{eq}$ is due to the increasing cooling rate when moving away from the intensity minimum of the cavity mode. 
However, as the particle is placed further away from the node, the cavity photon population builds up, and so does heating due to phase noise.
This behavior is quantitatively captured by our model [light-blue area behind the blue circles in Fig.~\ref{fig:fig3}(a), see Supplement~\cite{supp_mat_libration} for details], supporting further the hypothesis that our cooling performance is limited by phase noise.

\textit{Phase noise reduction.}
%
%
To combat phase noise in our experiment, we implement a phase-noise eater based on an imbalanced Mach-Zehnder interferometer for noise detection, and feedback to the laser via an electro-optic phase modulator~\cite{ivan2021}. The feedback gain $g$ determines the phase noise suppression, which we characterize in the Supplement~\cite{supp_mat_libration}. 
The remaining experiments in this article are carried out with a different nanoparticle, which has the same libration frequency $\Omega_\alpha$ as the one used for all experiments presented thus far~\cite{supp_mat_libration}. Furthermore, we use a homodyne libration detector cross-calibrated by sideband thermometry to boost detection efficiency to be able to resolve small occupation numbers~\cite{supp_mat_libration}.

Figure~\ref{fig:fig3}(a) shows as blue squares the occupation $n$ obtained with phase-noise cancellation gain $g=0.2$. The occupation reaches a minimum value of $n=0.25$ at a position $ky_{\mathrm{eq}}\approx0.2\pi$, closer to the cavity anti-node with respect to the situation without phase-noise cancellation. 
For gain $g=1$, shown as blue triangles in Fig.~\ref{fig:fig3}(a), the occupation monotonically decreases and reaches its minimum near the cavity anti-node ($ky_{\mathrm{eq}}\approx\pi/2$). This is the behavior expected in the absence of phase noise, where the optimum position of the particle for libration cooling is at the anti-node of the cavity field~\cite{stickler_elliptic_coherent}.
The lowest measured occupation is $n=0.04(1)$ quanta, corresponding to a ground-state purity of $92\%$. The blue areas in Fig.~\ref{fig:fig3}(a) are simultaneous theory fits to both data-sets taken with phase-noise cancellation. 
From these fits, we extract the shot noise heating rate (i.e., the heating rate in total absence of phase noise) $\Gamma_{\mathrm{BA}}/(2\pi)=\SI{0.5(1)}{\kilo\hertz}$.
We use this value to calculate the backaction-limited occupation that can be provided by our system, shown as the dashed black line in Figure~\ref{fig:fig3}(a). 
The proximity of the shot-noise limit to the data taken at cancellation gain $g=1$ (blue triangles) suggests that at this level of laser phase noise ($S_{\dot{\phi}\dot{\phi}}(\Omega_\alpha)=\SI{0.01}{\hertz^2/\hertz}$~\cite{supp_mat_libration}) the measured occupations are predominantly limited by quantum back-action. In Fig.~\ref{fig:fig3}(b), we present the homodyne spectra of the libration for the same gains used in Fig. \ref{fig:fig3}(a), with the particle positioned at the anti-node. The spectra were calibrated with sideband thermometry and are expressed in units of $\alpha_{\rm{zpf}}$, the zero-point angular displacement \cite{supp_mat_libration}.

Finally, we study the behavior of our system as a function of cancellation gain $g$ in some more detail. To this end, we set the detuning to its optimum value $\Delta=\Omega_\alpha$ and place the particle in the antinode of the cavity field.
Figure~\ref{fig:fig3}(c) shows the measured libration occupation $n$ as a function of cancellation gain $g$.
At $g=0$, despite otherwise optimal cooling conditions, excess phase noise leads to occupations above unity ($n>1$). As $g$ increases, the occupation decreases following our phase-noise model (blue area) and approaches the quantum back-action limit (black dashed line). 

In Fig.~\ref{fig:fig3}(d), we study the rates that determine the occupation of our levitated librator under cavity cooling as a function of cancellation gain $g$. 
First, we determine the optomechanical cooling rate $\gamma_{\mathrm{opt}}$ as the width of the Lorentzian fit to the libration peak, shown as light blue triangles in Fig.~\ref{fig:fig3}(d). We observe that $\gamma_{\mathrm{opt}}$ does not depend on cancellation gain $g$, as expected. 
Second, we extract the total heating rate $\Gamma$ of the libration as $\Gamma = \gamma_{\mathrm{opt}}\times n $~\cite{Delic2020}, and show it as red squares in Fig.~\ref{fig:fig3}(d). The total heating rate decreases with cancellation gain $g$ and approaches its fundamental limit $\Gamma_\mathrm{BA}$ (black dashed line). 
In the supplement, we present an independent measurement of the phase-noise heating rate performed by turning off the phase-noise cancellation and observing the population of the librator~\cite{supp_mat_libration}. 

\textit{Conclusions.} In conclusion, we have cooled a megahertz-frequency librational mode of an anisotropic levitated nanoparticle to the quantum ground state. 
Cooling was provided by coupling the photons  inelastically scattered by the particle to a high-finesse cavity in the resolved-sideband regime. 
We have used Raman sideband thermometry to determine the phonon occupation of the levitated librator.
Starting at room temperature, we have achieved a minimum occupation of 0.04(1) quanta, corresponding to a state purity of $92\%$.
A crucial technical step to reach this high purity was to minimize the impact of heating due to laser phase noise.
Active phase-noise cancellation in the tweezer beam by up to \SI{\sim20}{\decibel} put our system into a regime where the phonon occupation is a result of the balance between cavity cooling and heating by radiation torque shot noise, i.e., measurement backaction. 

The high purity achieved in our experiment places levitated librators on the forefront of experimental test-beds for room-temperature quantum optomechanics experiments~\cite{Huang_2024}. Furthermore, the purity of our system exceeds even that reached with giga-hertz frequency oscillators when laser-cooled in a cryogenic environment~\cite{Qiu_laserCooling-2020}.
The high-purity librational ground state could serve as a stepping stone towards preparing nonclassical states of motion~\cite{Stickler2021quantumrotations}.
An interesting first step would be to squeeze the high-purity librational state by modulation of the confining potential~\cite{rossi2024quantumdelocalizationlevitatednanoparticle}, in a free fall experiment~\cite{Stickler_2018_revivals}, or by exploiting unstable dynamics provided by the cavity~\cite{KusturaSqueezing_2022}.
Looking further, rotational motion exhibits genuine quantum effects with no counterparts in center-of-mass dynamics. Examples include orientational quantum revivals~\cite{Stickler_2018_revivals} and quantum-persistent tennis-racket flips~\cite{Ma2020tennis}.
Finally, the mega-hertz motional frequency demonstrated here may open up the possibility to resonantly couple levitated nanoparticles to other well-controlled quantum systems like trapped atomic ions~\cite{bykov2024nanoparticlestoredatomicion}, and to exploit qubit nonlinearities to engineer nonclassical states of motion~\cite{Najera2024_fluxonium}.

\textit{Acknowledgements.} 
We thank Titta Carlon Zambon, Fons van der Laan, Jialiang Gao,  Joanna Zielińska and the trappers of the Photonics Lab for stimulating discussions, and Ivan Galinsky, Vincent Dumont and Cédric Bugnon for help with the phase noise eater. C.G.B thanks Nadine Meyer for fruitful discussions. 
This research has been supported by the SERI Quantum Initiative (grant no. UeM019-2/SNF\_215917),  the SERI Quantum Call (grant no.  UeM029-3 / SNF\_225163) and by ETH Zurich (grant no. ETH-47 20-2).
L.D. acknowledges support from the Quantum Center Research Fellowship and the Dr Alfred and Flora Spälti Fonds. J.V. acknowledges support from the Dame Kathleen Ollerenshaw Fellowship of the University of Manchester.
CGB acklowledges support from the Austrian Science Fund (FWF) [10.55776/COE1].

\bibliography{bibliography}

\clearpage
\onecolumngrid
\pagenumbering{arabic}

\renewcommand{\thefigure}{S\arabic{figure}}
\renewcommand{\theequation}{S\arabic{equation}}
\renewcommand{\thesection}{S\arabic{section}}
\setcounter{equation}{0}
\setcounter{figure}{0}
\setcounter{section}{0}

\end{document}


\title{Librational ground-state cooling. Theory page for SM}


\section{Theoretical description}
\label{sec:theoretical_description}

We model the nanoparticle as a cylindrically symmetric dielectric particle whose polarizability tensor in the body frame $(\unitvec_X, \unitvec_Y, \unitvec_Z)$ takes the diagonal form
$
(\poltensorbf)_{ij}
=
\poleigenval_i
\delta_{ij}
$, where $i, j \in \lbrace X, Y, Z \rbrace$ and
$
\poleigenval_X
=
\poleigenval_Z
<
\poleigenval_Y
$.
The tweezer field, with frequency $\tweezfreq$, wavenumber $\tweezwavenumber$, power $\tweezpower$ and waist $\tweezwaist$, propagates along $\unitvec_z$ and is linearly polarized along the cavity axis ($\unitvec_y$).
In turn, the cavity mode has frequency $\cavfreq = \tweezfreq + \cavdetuning$, wavenumber $\cavwavenumber_c$, length $\cavlength$, waist $\cavwaist$ and a finite linewidth $\cavlinewidth$.
We place the origin of coordinates at the tweezer focus and assume the particle's center of mass is fixed at that position.
We also assume that the particle undergoes small-amplitude librations around its equilibrium orientation, which corresponds to its long axis being parallel to the tweezer field polarization
($\unitvec_Y
=
\unitvec_y$).
In particular, we focus on the small angular displacement $\angrot$ along $\unitvec_x$, leaving the one along $\unitvec_z$ as a straightforward generalization.
Assuming the particle is much smaller than all relevant optical wavelengths (point-dipole approximation), the Hamiltonian reads \cite{gonzalez-ballestero2023}
\begin{equation}
    \begin{aligned}
        \ham (t)
        =
        \frac{\angmom^2}
        {2 \mominertia}
        &
        +
        \hbar
        \cavfreq
        \cavcre
        \cavanh
        +
        \hamfree
        \\
        &
        -
        \frac{1}{2}
        \electricfieldvec
        (\mathbf{0}, t)
        \poltensorlf
        (\angrot)
        \electricfieldvec
        (\mathbf{0}, t)
        +
        \hamdrive (t).
        \label{eq:initial_hamiltonian}
    \end{aligned}
\end{equation}
Here, $\angmom$ and $\mominertia$ are angular momentum and moment of inertia of the nanoparticle, $\cavanh$, $\cavcre$ are the ladder operators of the cavity mode, and $\hamfree$ is the Hamiltonian of the free-space electromagnetic modes \cite{gonzalez-ballestero2019}.
The fourth term in Eq. \eqref{eq:initial_hamiltonian} describes the optomechanical interaction, where $\poltensorlf (\angrot)$ is the polarizability tensor of the particle in the laboratory frame $(\unitvec_x, \unitvec_y, \unitvec_z)$ and
$
\electricfieldvec (\mathbf{0}, t)
=
\tweezfieldvec (\mathbf{0}, t)
+
\cavfieldvec (\mathbf{0})
+
\freefieldvec (\mathbf{0})
$
the total electric field at the position of the particle. This consists of (i) the classical tweezer field
$
\tweezfieldvec (\mathbf{0}, t)
=
(\unitvec_y /2)
[
\tweezfieldampl
e^{\imconst \phasenoise (t)}
e^{\imconst \tweezfreq t}
+
\text{c.c.}
]
$,
where
$
\tweezfieldampl
=
\sqrt{
4 \tweezpower
/
(\pi
\dielpermvacuum
\speedlight
\tweezwaist^2
)
}
$
is the field amplitude, with
$\dielpermvacuum$ the dielectric permittivity of vacuum and $\speedlight$ the speed of light, and
$e^{\imconst \phasenoise (t)}$
is a noisy phase factor whose properties we discuss below;
(ii) the cavity field
$
\cavfieldvec (\mathbf{0})
=
\unitvec_x
\cavfieldampl
\sin (\cavphase)
(
\cavcre
+
\cavanh
)
$,
where $\cavphase$ encodes the position of the particle along the cavity intensity profile ($\cavphase = 0$ at the node, $\cavphase = \pi/2$ at the antinode), and
$
\cavfieldampl
=
\sqrt{
\hbar
\cavfreq
/
(2 \dielpermvacuum
\cavvolume)
}
$ is the zero-point cavity field amplitude,
with
$
\cavvolume
=
\pi
(\cavwaist/2)^2
\cavlength
$
the mode volume of the cavity;
(iii) the free-space field $\freefieldvec (\mathbf{0})$, given in Ref. \cite[Eq. (7)]{gonzalez-ballestero2019}.
Lastly, the driving term
$\hamdrive (t) = -
\hbar
\cavdrivingfreq
(
\cavcre
+
\cavanh
)
(
e^{\imconst \tweezfreq t}
e^{\imconst \phasenoise (t)}
+
\text{c.c.}
)$
models the Rayleigh scattering of tweezer photons into the cavity, an effect that is due to experimentally uncontrolled deviations from the ideal cooling conditions (\textit{e.g.} imperfect polarization alignment).
The cavity driving rate reads
$
\cavdrivingfreq
=
\poleigenval_Y
\tweezfieldampl
\cavfieldampl
\cavdrivingconst
\sin (\cavphase)
/ 
2 \hbar
$,
where we introduced the small adimensional parameter $\cavdrivingconst$. 

The statistical properties of $e^{\imconst \phasenoise (t)}$ derive from those of the Gaussian white-noise process $\dot{\phasenoise} (t)$, which has first and second order moments
$
\llangle
\dot{\phasenoise} (t)
\rrangle
=
0
$,
$
\llangle
\dot{\phasenoise} (t)
\dot{\phasenoise} (s)
\rrangle
=
\phasenoisepsd
\delta (t - s)
$.
Here,
$
\llangle
\sbullet
\rrangle
$
indicates the average over all noise realizations and $\phasenoisepsd$ is the noise power spectral density (PSD).

Assuming the particle is close to its equilibrium orientation, we proceed as follows. (1) We expand the optomechanical interaction in the limit of small librations.
In doing so, we neglect three kinds of terms:
(i) a term proportional to
$
\libpos^2
(\cavcre
+
\cavanh)^2
$,
where
$
\libpos
=
\libcre
+
\libanh
$
and $\libanh$, $\libcre$ are the ladder operators of the librational mode, since it is subdominant for the numbers of this experiment,
(ii) two nanoparticle-mediated coupling terms between cavity and free-space modes, which can be neglected as otherwise they would lead to a renormalization of the cavity linewidth which is not observed for subwavelength particles, and
(iii) two terms shifting the energy of the free-space electromagnetic modes, an effect again negligible for subwavelength particles.
(2) We transform the resulting linear Hamiltonian to a frame rotating at the tweezer frequency $\tweezfreq$ and we undertake the rotating wave approximation (RWA) to remove terms oscillating at $2 \tweezfreq$ \cite{gonzalez-ballestero2019}.
(3) We trace out the free-space electromagnetic modes in the Born-Markov approximation, obtaining a reduced master equation for the particle-cavity dynamics.
(4) We apply a displacement transformation $\hat{c} \to \hat{c} + \alpha_c(t)$, where  $\cavdispl (t)$ is the classical cavity amplitude under the stochastic driving induced by phase noise, \textit{i.e.}
$
\dot{\cavdispl} (t)
=
- 
\left(
\imconst
\cavdetuning
+
\cavlinewidth/2
\right)
\cavdispl (t)
+
\imconst
\cavdrivingfreq
e^{- \imconst \phasenoise (t)}
$.
At this stage, the master equation for the density operator of cavity and librational modes, $\cavlibstate$, reads
\begin{equation}
    \dot{\cavlibstate} (t)
    =
    -\frac{\imconst}{\hbar}
    [
    \hamdispl (t),
    \cavlibstate (t)
    ]
    +
    \frac{\cavlinewidth}{2}
    \lindbladian_{\cavanh}
    [\cavlibstate (t)]
    -
    \frac{\recoilheatrate}{2}
    [
    \libpos,
    [
    \libpos,
    \cavlibstate (t)
    ]
    ].
    \label{eq:master_equation_after_displacement}
\end{equation}
The Hamiltonian reads
\begin{equation}
    \hamdispl (t)/\hbar
    =
    \libfreq
    \libcre
    \libanh
    +
    \mechdriving (t)
    \libpos
    +
    \cavdetuning
    \cavcre
    \cavanh
    +
    \couplingpartcav
    \libpos
    (
    \cavanh
    e^{\imconst \phasenoise (t)}
    +
    \cavcre
    e^{-\imconst \phasenoise (t)}
    ),
    \label{eq:hamiltonian_after_displacement}
\end{equation}
where we introduced the librational frequency
$
\libfreq
=
\sqrt{
(\poleigenvaldiff)
\tweezfieldampl^2
/
2 \mominertia
}
$,
the coupling strength
$
\couplingpartcav
=
-
(\poleigenvaldiff)
\libzpm
\tweezfieldampl
\cavfieldampl
\sin (\cavphase)
/
2 \hbar
$
and the stochastic mechanical driving
$
\mechdriving (t)
=
2
\couplingpartcav
\Re
\left[
e^{\imconst \phasenoise (t)}
\cavdispl (t)
\right]
$,
as well as
$
\poleigenvaldiff
=
\poleigenval_Y
-
\poleigenval_X
$
and the mechanical zero-point angular displacement
$
\libzpm
=
\sqrt{
\hbar
/ (2 \mominertia
\libfreq)
}
$.
Eq. \eqref{eq:master_equation_after_displacement} also includes a Lindblad dissipator for the decay of the cavity mode
($
\lindbladian_{\hat{a}} [\cavlibstate]
=
\hat{a}
\cavlibstate
\hat{a}^\dagger
-
\frac{1}{2}
\lbrace
\hat{a}^\dagger
\hat{a},
\cavlibstate
\rbrace
$)
and a position-localization dissipator representing photon recoil heating (shot noise), with rate
\begin{equation}
    \recoilheatrate
    =
    \frac{
    (\poleigenvaldiff)^2
    \tweezfieldampl^2
    \libzpm^2
    \tweezfreq^3
    }{
    (12 \pi \hbar c^3
    \dielpermvacuum)
    }.
    \label{eq:recoil_heating_rate}
\end{equation}
One can show that the noisy multiplicative phase factor in the last term of Eq. \eqref{eq:hamiltonian_after_displacement} is negligible, $e^{i\varphi(t)}\approx 1$, under the assumption
$
\phasenoisepsd
\ll
\cavlinewidth
$,
which is the regime of our experiment.
Hence, the main effect of phase noise is a stochastic force proportional to $\mechdriving (t)$, which leads to additional heating of the particle's motion.

To explore the librational dynamics we adiabatically eliminate the cavity mode assuming
$
\couplingpartcav
\ll
\cavlinewidth
$
\cite{gonzalez-ballestero2023tutorial, wilson-rae2008} and we average over all possible realizations of the noise following the second-order generalized cumulant expansion method \cite{vankampen1974a, vankampen1974b}.
After applying a final displacement transformation
$
\libanh
\rightarrow
\libanh
+
\lim_{t \rightarrow \infty}
\llangle
\mechdriving (t)
\rrangle
$, the master equation for the density operator of the librations, $\libstate$, reads
\begin{equation}
    \begin{aligned}
        \dot{\libstate} (t)
        =
        -\imconst
        [
        \libfreq
        \libcre
        \libanh,
        \libstate (t)
        ]
        &
        -
        \frac{1}{2}
        (
        \recoilheatrate
        +
        \phasenoiseheatrate
        )
        [
        \libpos,
        [
        \libpos,
        \libstate (t)
        ]
        ]
        \\
        &
        +
        \aecrate_+
        \lindbladian_{\libcre}
        [\libstate (t)]
        +
        \aecrate_-
        \lindbladian_{\libanh}
        [\libstate (t)].
        \label{eq:final_master_equation}
    \end{aligned}
\end{equation}
The adiabatic elimination of the cavity leads to the Lindbladians for heating and cooling in the second line, with rates
$
\aecrate_{\pm}
=
\couplingpartcav^2
\cavlinewidth
/
[(\libfreq \pm \cavdetuning)^2
+
\left(
\cavlinewidth/2
\right)^2]
$,
as well as a small shift of the libration frequency which can be neglected for this experiment.
Since in our experiment
$\aecrate_+ \ll \aecrate_-$,
the total cooling rate
$
\totcoolrate
=
\aecrate_+
-
\aecrate_-
$
can be approximated as
\begin{equation}
    \totcoolrate
    \approx
    -\aecrate_-
    =
    -
    \frac{
    \couplingpartcav^2
    \cavlinewidth
    }{
    (\libfreq - \cavdetuning)^2
    +
    \left(
    \frac{\cavlinewidth}{2}
    \right)^2}.
    \label{eq:total_cooling_rate}
\end{equation}
Note that phase noise induces an additional heating term with rate $\Gamma_\varphi$, which in the weak noise limit
$
\phasenoisepsd
\ll
\cavlinewidth
$
reads
\begin{equation}
    \phasenoiseheatrate
    \approx
    \frac{4
    \couplingpartcav^2
    \ncavss
    \phasenoisepsd
    \lbrace
    [
    \left(
    \frac{\cavlinewidth}{2}
    \right)^2
    -
    \cavdetuning^2
    ]^2
    +
    \left(
    \libfreq
    \frac{\cavlinewidth}{2}
    \right)^2
    \rbrace
    }
    {
    [
    \cavdetuning^2
    +
    \left(
    \frac{\cavlinewidth}{2}
    \right)^2
    ]
    [
    (
    \cavdetuning
    +
    \libfreq
    )^2
    +
    \left(
    \frac{\cavlinewidth}{2}
    \right)^2
    ]
    [
    (
    \cavdetuning
    -
    \libfreq
    )^2
    +
    \left(
    \frac{\cavlinewidth}{2}
    \right)^2
    ]
    },
    \label{eq:phase_noise_heating_rate_weak_noise}
\end{equation}
where
$
\ncavss
=
\lim_{t \rightarrow \infty}
\llangle
\lvert
\cavdispl
(t)
\rvert^2
\rrangle
\approx
\cavdrivingfreq^2
/
[
\cavdetuning^2
+
\left(
\cavlinewidth / 2
\right)^2
]
$
is the steady-state cavity occupation in the weak noise limit.
In Eq. \eqref{eq:final_master_equation} we neglect any dissipation due to collisions with gas molecules, as they become negligible at the experiment's pressure
$
\gaspressure
=
5
\times
10^{-9}
$
mbar \cite{van-der-laan2021}.

Given Eq. \eqref{eq:final_master_equation} we can compute the steady-state number of librational phonons as
$
\nphononsss
=
\lim_{t \rightarrow \infty}
\tr
\lbrace
\libcre \libanh
\libstate (t)
\rbrace
=
-(\recoilheatrate
+
\phasenoiseheatrate)
/
\totcoolrate
$, where we neglected the cavity contribution to the total heating rate since
$
\aecrate_+
\ll
\recoilheatrate
$.
This leads to the expression
\begin{widetext}
\begin{equation}
    \begin{aligned}
        \nphononsss
        =
        \nphononsssnot
        +
        \nphononsssphasenoise
        =
        \frac{
        \recoilheatrate
        [
        (
        \cavdetuning
        -
        \libfreq
        )^2
        +
        \left(
        \frac{\cavlinewidth}{2}
        \right)^2
        ]
        }{
        \couplingpartcav^2
        \cavlinewidth}
        +
        \frac{4
        \ncavss
        \phasenoisepsd
        \lbrace
        [
        \left(
        \frac{\cavlinewidth}{2}
        \right)^2
        -
        \cavdetuning^2
        ]^2
        +
        \left(
        \libfreq
        \frac{\cavlinewidth}{2}
        \right)^2
        \rbrace
        }
        {
        [
        \cavdetuning^2
        +
        \left(
        \frac{\cavlinewidth}{2}
        \right)^2
        ]
        [
        (
        \cavdetuning
        +
        \libfreq
        )^2
        +
        \left(
        \frac{\cavlinewidth}{2}
        \right)^2
        ]
        \cavlinewidth
        },
    \end{aligned}
    \label{eq:phonon_number_steady_state}
\end{equation}
\end{widetext}
which has been used to fit the experimental data.
The two terms on the right-hand side of Eq. \eqref{eq:phonon_number_steady_state} are respectively the phonon number in the absence of phase noise, $\nphononsssnot$, and the contribution due to phase noise, $\nphononsssphasenoise$.
In the regime of weak noise
($
\phasenoisepsd
\ll
\cavlinewidth
$),
resolved sidebands
($
\cavlinewidth
\ll
\libfreq
$)
and optimal detuning
($
\cavdetuning
=
\libfreq
$),
$\nphononsssphasenoise$ takes the approximate form
$
\nphononsssphasenoise
\approx
\ncavss
\phasenoisepsd
/
\cavlinewidth
$,
in agreement with the expressions known in literature \cite{rabl2009, delic2019, meyer2019}.

\bibliography{SM_theory_bibliography}